# On the derivation of the equilibrium equations in terms of displacement

Peng SHI[1, 2] *

[1] Logging Technology Research Institute, China National Logging Corporation, No.50 Zhangba five road, Xi'an, 710077, P. R. China

[2] Well Logging Technology Pilot Test Center, China National Logging Corporation, No.50 Zhangba five road, Xi'an, 710077, P. R. China

*Corresponding author: sp198911@outlook.com;

***Abstract.*** The study shows that errors exist in the derivation of equilibrium equations in terms of displacement. It is discovered that when the equilibrium equations in terms of displacement are derived, the variation of the differential order of displacement may cause the variation of the stress state in an elastomer. For plane stress problems, the Lame-Navier equations are not equivalent to the equilibrium equations described with stress. By submitting the displacement field of the well-known issue of a rectangular beam purely bent into the Lame-Navier equations, the conclusion is confirmed. It is also revealed that the velocity of longitudinal wave in an elastomer is affected by its thickness.

# 1. Introduction

The theory of elasticity studies the stress state and deformation of elastomer under various loads. The theory is continuously developed and improved in the process of solving practical engineering problems. Galileo first looked into the bending problem of beams in 1638, driven by the demands of building projects [1, 2]. In 1678, Hooke published the physical law, now known as Hooke's law. It states that the deformation of an elastic body is proportional to the applied force based on experimental results of metal wires, springs, and cantilever beams [1, 2]. The general equation of the theory of elasticity was derived by Navier and Cauchy in 1821 and 1822, providing the theoretical framework for elasticity.

In the theory of elasticity, elastomers are considered to be perfectly continuous and are paid no attention to their molecular structure, which is called continuum hypothesis [3, 4]. The motion of material elements constituting an elastomer is believed to be the motion of particles, which can be described by Newton's three laws of motion or other mechanical principles related to and equivalent to them [1-7]. Continuum hypothesis allows for the description of internal force acting on every given surface element in the form of a field and the use of powerful methods of calculus to describe the equilibrium of a free body with an infinitesimal volume in an elastomer [3, 4]. In order to conveniently describe the force on the bounding surface of a free body and the equilibrium of the free body whose volume goes to zero under resultant force, the stress tensor is introduced into the theory of elasticity, which is a second-order symmetric tensor [5, 6]. Since stress are caused by strain in an elastomer, the strain tensor is introduced to describe the constitutive relation of an elastomer and the deformation at a certain point.

Many analytical techniques, such as the displacement solution method, stress-based solution method, stress function method, etc., have been proposed for solving simple elastic problems [1, 7-9]. The displacement solution method is the method that takes displacement as an unknown variable. The stress-based solution method is the method that takes the stress component as an unknown variable. The stress function method is the method by introducing functions that that automatically fulfill the equilibrium equations. Although different solution methods use different variables and functions, due to the uniqueness of the solutions for the same elastic problem, the stress and displacement fields obtained by different solution methods should be the same. The Lame-Navier equations are believed to be the equilibrium equations of an elastomer in terms of displacement, and are the basic equations of displacement solution method. However, my recent efforts on the theory of elasticity indicate that the displacement fields obtained from stress function method may not satisfy the Lame-Navier equations.

The study aims to prove that the Lame-Navier equations are not the equilibrium equations in terms of displacement. By rederiving the equilibrium equations in terms of displacement, it is revealed that the variation of the differential order of displacement in the derivation of the Lame-Navier equations causes the variation of stress state in an elastomer. The result is verified by putting the displacement fields of the well-known problem of a rectangular beam purely bent into the Lame-Navier equations.

## 2. Derivation of the equilibrium equations in terms of displacement

In the Cartesian coordinate system, the equilibrium equations in the form of components are expressed as follows:

$$\begin{cases} \dfrac{\partial \sigma_{xx}}{\partial x} + \dfrac{\partial \sigma_{xy}}{\partial y} + \dfrac{\partial \sigma_{xz}}{\partial z} = \rho \dfrac{\partial^2 u}{\partial t^2} \\ \dfrac{\partial \sigma_{yx}}{\partial x} + \dfrac{\partial \sigma_{yy}}{\partial y} + \dfrac{\partial \sigma_{yz}}{\partial z} = \rho \dfrac{\partial^2 v}{\partial t^2} \\ \dfrac{\partial \sigma_{zx}}{\partial x} + \dfrac{\partial \sigma_{zy}}{\partial y} + \dfrac{\partial \sigma_{zz}}{\partial z} = \rho \dfrac{\partial^2 w}{\partial t^2} \end{cases} \quad (1)$$

where, $\sigma_{ij}$ is the stress component ($i, j = x, y, z$), $\rho$ is the density of elastomer, $u$, $v$ and $w$ are the displacement component along $x$-axis, $y$-axis and $z$-axis, respectively. For an isotropic elastomer, stress component can be expressed with displacement as follows [1, 5, 6-9]:

$$\begin{cases} \sigma_{xx} = \lambda \theta + 2\mu \dfrac{\partial u}{\partial x} \\ \sigma_{yy} = \lambda \theta + 2\mu \dfrac{\partial v}{\partial y} \\ \sigma_{yy} = \lambda \theta + 2\mu \dfrac{\partial w}{\partial z} \\ \sigma_{xy} = \sigma_{yx} = \mu \left( \dfrac{\partial u}{\partial y} + \dfrac{\partial v}{\partial x} \right) \\ \sigma_{xz} = \sigma_{zx} = \mu \left( \dfrac{\partial u}{\partial z} + \dfrac{\partial w}{\partial x} \right) \\ \sigma_{yz} = \sigma_{zy} = \mu \left( \dfrac{\partial v}{\partial z} + \dfrac{\partial w}{\partial y} \right) \end{cases} \quad (2)$$

where, $\lambda$ and $\mu$ are lame coefficient, and $\theta = \partial u/\partial x + \partial v/\partial y + \partial w/\partial z$. Substituting Equations (2) into Equations (1), the equilibrium equations in terms of displacement are obtained.

Substituting Equation (2) into Equation (1), the $x$-component of the equilibrium

equation in terms of displacement is expressed as:

$$\lambda\frac{\partial\theta}{\partial x}+2\mu\frac{\partial^2 u}{\partial^2 x}+\mu\frac{\partial}{\partial y}\left(\frac{\partial u}{\partial y}+\frac{\partial v}{\partial x}\right)+\mu\frac{\partial}{\partial z}\left(\frac{\partial u}{\partial z}+\frac{\partial w}{\partial x}\right)=\rho\frac{\partial^2 u}{\partial t^2} \quad (3)$$

In Equation (3), The first two terms represent the differential of $\sigma_{xx}$ with respect to $x$, the third term represents the differential of $\sigma_{xy}$ with respect to $y$, and the fourth term represents the differential of $\sigma_{xz}$ with respect to $z$. In traditional derivation, by changing the differential order of displacement, the $x$ and $y$ components of the equilibrium equations in terms of displacement are rewritten as [1, 6]:

$$\begin{cases}(\lambda+\mu)\dfrac{\partial}{\partial x}\left(\dfrac{\partial u}{\partial x}+\dfrac{\partial v}{\partial y}+\dfrac{\partial w}{\partial z}\right)+\mu\left(\dfrac{\partial^2}{\partial^2 x}+\dfrac{\partial^2}{\partial^2 y}+\dfrac{\partial^2}{\partial^2 z}\right)u=\rho\dfrac{\partial^2 u}{\partial t^2}\\ (\lambda+\mu)\dfrac{\partial}{\partial y}\left(\dfrac{\partial u}{\partial x}+\dfrac{\partial v}{\partial y}+\dfrac{\partial w}{\partial z}\right)+\mu\left(\dfrac{\partial^2}{\partial^2 x}+\dfrac{\partial^2}{\partial^2 y}+\dfrac{\partial^2}{\partial^2 z}\right)v=\rho\dfrac{\partial^2 v}{\partial t^2}\end{cases} \quad (4)$$

Equations (4) are $x$ and $y$ components of the Lame-Navier equations. It should be pointed out that Equations (4) do not always hold true.

Assuming that a research problem is a plane stress problem in the plane of ($xoy$), $\sigma_{zz}$, $\sigma_{xz}$, and $\sigma_{yz}$ in Equations (1) are zero. In this case, the equilibrium equations in terms of displacement are expressed as:

$$\begin{cases}\lambda\dfrac{\partial}{\partial x}\left(\dfrac{\partial u}{\partial x}+\dfrac{\partial v}{\partial y}+\dfrac{\partial w}{\partial z}\right)+2\mu\dfrac{\partial^2 u}{\partial^2 x}+\mu\dfrac{\partial}{\partial y}\left(\dfrac{\partial u}{\partial y}+\dfrac{\partial v}{\partial x}\right)=\rho\dfrac{\partial^2 u}{\partial t^2}\\ \mu\dfrac{\partial}{\partial x}\left(\dfrac{\partial u}{\partial y}+\dfrac{\partial v}{\partial x}\right)+\lambda\dfrac{\partial}{\partial y}\left(\dfrac{\partial u}{\partial x}+\dfrac{\partial v}{\partial y}+\dfrac{\partial w}{\partial z}\right)+2\mu\dfrac{\partial^2 v}{\partial^2 y}=\rho\dfrac{\partial^2 v}{\partial t^2}\end{cases} \quad (5)$$

If we directly obtain the Lame-Navier equations of a plane stress problem through degenerating Lame-Navier equations, the following Equations are obtained:

$$\begin{cases}(\lambda+\mu)\dfrac{\partial}{\partial x}\left(\dfrac{\partial u}{\partial x}+\dfrac{\partial v}{\partial y}+\dfrac{\partial w}{\partial z}\right)+\mu\left(\dfrac{\partial^2}{\partial^2 x}+\dfrac{\partial^2}{\partial^2 y}\right)u=\rho\dfrac{\partial^2 u}{\partial t^2}\\ (\lambda+\mu)\dfrac{\partial}{\partial y}\left(\dfrac{\partial u}{\partial x}+\dfrac{\partial v}{\partial y}+\dfrac{\partial w}{\partial z}\right)+\mu\left(\dfrac{\partial^2}{\partial^2 x}+\dfrac{\partial^2}{\partial^2 y}\right)v=\rho\dfrac{\partial^2 v}{\partial t^2}\end{cases} \quad (6)$$

Equations (5) and Equations (6) are different. This means that the Lame-Navier equations are not applicable to plane stress problems. Comparing Equations (5) and Equations (6), it is obtained that the special constitutive relationship of elastomers that strain does not necessarily generate stress is the reason why the differential order of displacement cannot be changed arbitrarily during the derivation of equilibrium equations in terms of displacement.

For a plane stress problem, the following relationship holds [1, 8, 9]:

$$\frac{\partial w}{\partial z} = -\frac{\upsilon}{1-\upsilon}\left(\frac{\partial u}{\partial x} + \frac{\partial v}{\partial y}\right) \quad (7)$$

with $\upsilon$ the Poisson's ratio. Submitting Equation (7) into Equations (5), the equilibrium equations in terms of displacement for plane stress problems are rewritten as:

$$\begin{cases} \lambda \frac{1-2\upsilon}{1-\upsilon}\frac{\partial}{\partial x}\left(\frac{\partial u}{\partial x} + \frac{\partial v}{\partial y}\right) + 2\mu\frac{\partial^2 u}{\partial^2 x} + \mu\frac{\partial}{\partial y}\left(\frac{\partial u}{\partial y} + \frac{\partial v}{\partial x}\right) = \rho\frac{\partial^2 u}{\partial t^2} \\ \mu\frac{\partial}{\partial x}\left(\frac{\partial u}{\partial y} + \frac{\partial v}{\partial x}\right) + \lambda\frac{1-2\upsilon}{1-\upsilon}\frac{\partial}{\partial y}\left(\frac{\partial u}{\partial x} + \frac{\partial v}{\partial y}\right) + 2\mu\frac{\partial^2 v}{\partial^2 y} = \rho\frac{\partial^2 v}{\partial t^2} \end{cases} \quad (8)$$

Equations (8) indicate that the velocity of longitudinal waves propagating in an elastic plate increases with the plate thickness. With Equation (8), the velocity of plane longitudinal wave in a flat plate, which can be regarded as a plane stress problem, is expressed as:

$$c = \sqrt{\frac{(1-2\upsilon)\lambda}{(1-\upsilon)\rho} + \frac{2\mu}{\rho}} \quad (9)$$

Further considering a thin rod, when its cross-section is small enough, the stress during longitudinal wave propagation can be considered as uniaxial tension [10]. In this case,

the longitudinal wave velocity will approach the minimum value:

$$c = \sqrt{\frac{E}{\rho}} \tag{10}$$

The velocity is the same as the velocity of longitudinal wave in rod theory [10]. The result further support that errors exist in the traditional derivation of equilibrium equations in terms of displacement.

**3. Case study**

Considering a rectangular beam having a narrow cross section of unit width bent under a constant bending moment (Figure 1), the upper and lower edges are free from loads, and the width of rectangular beam is much smaller than its height. In this case, the bending of the rectangular beam can be regarded as a plane stress problem. Assuming the stress function for pure bending of rectangular beam is as follows [1]:

$$\phi = \frac{M}{4c^3} y^3 \tag{11}$$

with $M$ the constant bending moment and $c$ the half height of rectangular beam, the stress components are obtained:

$$\begin{cases} \sigma_{xx} = \frac{3M}{2c^3} y \\ \sigma_{yy} = 0 \\ \sigma_{xy} = 0 \end{cases} \tag{12}$$

For the rectangular beam, the displacement components are obtained by submitting Equations (2) into Equations (12) [1, 8]:

$$\begin{cases} u = \dfrac{M}{EI}xy - \omega y + u_0 \\ v = -\dfrac{\upsilon M}{2EI}y^2 - \dfrac{M}{2EI}x^2 + \omega x + w_0 \end{cases} \quad (13)$$

where, $I=2c^3/3$, $\omega$, $u_0$ and $w_0$ are the integration constants.

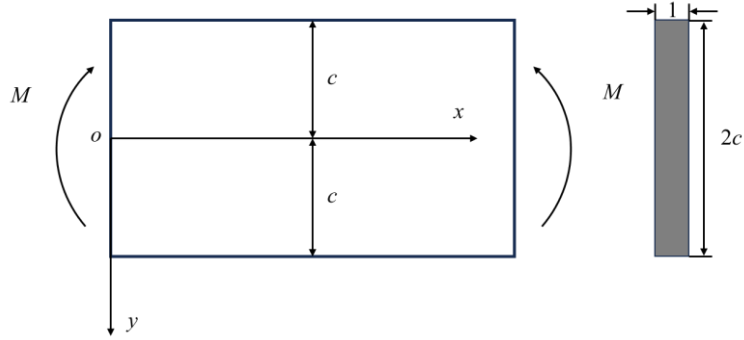

**Figure 1. Sketch of the pure bending of rectangular beam**

Since Equations (13) are the displacement obtained by submitting Equations (2) into Equations (12), the stress component described by displacement should be the same as Equations (12), and Equations (8) should be automatically satisfied. However, submitting Equations (7) and (13) into Equations (6), the following relationships are obtained:

$$\begin{cases} (\lambda+\mu)\dfrac{\partial}{\partial x}\left(\dfrac{\partial u}{\partial x}+\dfrac{\partial v}{\partial y}+\dfrac{\partial w}{\partial z}\right)+\mu\left(\dfrac{\partial^2}{\partial^2 x}+\dfrac{\partial^2}{\partial^2 y}\right)u = 0 \\ (\lambda+\mu)\dfrac{\partial}{\partial y}\left(\dfrac{\partial u}{\partial x}+\dfrac{\partial v}{\partial y}+\dfrac{\partial w}{\partial z}\right)+\mu\left(\dfrac{\partial^2}{\partial^2 x}+\dfrac{\partial^2}{\partial^2 y}\right)v = -\dfrac{\upsilon\mu M}{EI} \end{cases} \quad (14)$$

Equations (14) indicates that the Lame-Navier equations are not equivalent to the equilibrium equations described with stress.

## 4. Conclusions

The study has shown that the Lame-Navier equations are not equivalent to the equilibrium equations described with stress. Errors occur when the equilibrium equations in terms of displacement are derived. The special constitutive relationship of elastomers that strain does not necessarily generate stress require that the differential order of displacement cannot be changed arbitrarily during the derivation of equilibrium equations in terms of displacement. The conclusion is confirmed by substituting the displacement field of the well-known problem of a rectangular beam purely bent into Lame-Navier equations. The equilibrium equations in terms of displacement implies that the velocity of longitudinal wave propagating in an elastomer is influenced by the thickness of elastomer.

**Acknowledgments**

This work was supported by the scientific research and technology development project of China National Petroleum Corporation (2020B-3713).

**Declaration of competing interest**

The author declares that he has no known competing financial interests or personal relationships that could have appeared to influence the work reported in this paper.